\documentclass[twocolumn,aps,prl,superscriptaddress,nofootinbib,showpacs,floatfix]{revtex4}

\usepackage{multirow}
\usepackage{epsfig}
\usepackage{amsmath}
\usepackage[normalem]{ulem}  % \sout{old text} for strikeout
\usepackage[dvips]{color} % For blue in-text comments and additions

\renewcommand\sout{\bgroup \color{red} \ULdepth=-.5ex \ULset}
\newcommand{\Ex}[2]{\ifmmode{#1\times10^{#2}}\else{$#1\times10^{#2}$}\fi}
\begin{document}
\title{Free energy versus internal energy potential for heavy quark systems at finite temperature}

\author{Su Houng~Lee}\affiliation{Department of Physics and Institute of Physics and Applied Physics, Yonsei
University, Seoul 120-749, Korea}
\author{Kenji Morita}\affiliation{Yukawa Institute for Theoretical
Physics, Kyoto University, Kyoto 606-8502, Japan}
%\affiliation{RIKEN Nishina Center, Hirosawa 2-1, Wako, Saitama 351-0198, Japan}
\author{Taesoo Song}\affiliation{Cyclotron Institute and  Department of Physics and Astronomy, Texas A\&M
University, College Station, Texas 77843, U.S.A.}
\author{Che Ming Ko}\affiliation{Cyclotron Institute and  Department of Physics and Astronomy, Texas A\&M
University, College Station, Texas 77843, U.S.A.}

\date{\today}
\begin{abstract}
Using the QCD sum rule with its operator product expansion reliably
estimated from lattice calculations for the pressure and energy density
of hot QCD matter, we calculate the strength of the
$J/\psi$ wave function at  origin and find that it decreases with
temperature when the temperature is above the transition temperature. This result is shown to
follow exactly that obtained from the solution of the Schr\"odingier
equation for a charm and anticharm quark pair using the free energy
from lattice calculations as the potential and is in sharp
contrast to that using the deeper potential associated with the
internal energy, which shows an enhanced strength of the $J/\psi$ wave
function at  origin. Our result thus has resolved  the long-standing
question of whether the free energy potential or the internal energy
potential should be used in analyzing the spectrum of heavy quark
systems at finite temperature.
\end{abstract}


\maketitle

Ever since the seminal work by Matsui and Satz~\cite{Matsui:1986dk},
suggesting that the $J/\psi$ suppression could be a signature for the
formation of the quark-gluon plasma (QGP) in relativistic heavy ion
collisions~\cite{Scomparin:2012vq}, there have been numerous theoretical
as well as experimental studies from SPS to LHC on this
subject.  On the other hand, it is quite discomforting to know that even
to this date, the question whether $J/\psi$  dissolves immediately
above the critical temperature $T_c~$\cite{Mocsy:2007jz,Gubler:2011ua,Ding:2012sp} or remains
bounded at higher temperatures~\cite{Asakawa:2003re,Datta:2003ww,Ohno:2011zc} is
still not settled.   The two scenarios have led to different
phenomenological models based on either complete melting or sequential suppressions of charmonia in QGP for  explaining the observed suppression of $J/\psi$ production in relativistic heavy ion collisions, resulting  thus in different conclusions about the phase transition in QCD and the properties of the quark-gluon plasma~\cite{Song:2011xi}.

Given the fact that lattice calculations can calculate the heavy quark
potential with great precision, it might seem that the question could
be simply answered by just solving the Schr\"{o}dinger equation and
finding the ground state eigenvalues.  However, it is not so clear
whether one should use the free energy potential or the internal energy
potential~\cite{Kaczmarek:1900zz}. If one uses the free energy as the
potential, the $J/\psi$ then dissolves slightly above $T_c$ whereas  it
remains bounded up to almost $2T_c$ if the internal energy is used. The
difference between the two potentials can be traced to the question of
whether one should subtract out the additional gluonic entropy
contribution associated with adding the heavy quark pair in the
QGP~\cite{Wong:2004zr}. Because of this difference, the internal energy
potential is deeper and also rises more sharply with increasing  
interquark distance, resulting in a $J/\psi$ ground state wave function
that is more localized at  origin than that from the free energy
potential. This effect leads to marked different behaviors in the
temperature dependence of the $J/\psi$ wave function at  origin.

The $J/\psi$ wave function at origin is the non-relativistic limit of the
overlap of the $J/\psi$ with the charm vector current and is thus a well
defined gauge invariant and physically observable quantity.  In this
work, using QCD sum rules at finite temperature, we show that the
temperature dependence of this overlap follows exactly the temperature
dependence of the $J/\psi$ wave function at  origin that is obtained from
using the free energy potential in the Schr\"{o}dinger equation for
the charm and anticharm quark pair.  Unlike the uncertainty of the QCD
sum rule approaches in predicting particular changes in the whole
spectral density, the overlap can be obtained from the integrated part
of the spectral density and thus reliably estimated.  In fact, we will show
that the zero temperature sum rule reproduces the strength needed
to explain the dielectron partial width of the $J/\psi$ and that the finite temperature calculations are equally
reliable.

We start with the correlation function of the charm vector current
$J_\mu=\bar{c}\gamma_\mu c$,
\begin{eqnarray}
\Pi_{\mu \nu}(q) & = & i\int d^4x \,e^{iqx}\langle T[ J_\mu(x) J_\nu(0) ] \rangle.
\end{eqnarray}
Inserting $J/\psi$ as an intermediate state using the normalization $
\langle J/\psi (p) |J/\psi (p) \rangle =(2 \pi)^3 2\omega_p \delta^3(p-p)$,
we have
\begin{eqnarray}
\Pi_{\mu}^{\mu}(q) & = & - \frac{ \langle J_\mu(0)|J/\psi \rangle
 \langle J/\psi | J^\mu(0)\rangle }{q^2-m_{J/\psi}^2} ,
\end{eqnarray}
where
\begin{eqnarray}
\langle J_\mu (0) |J/\psi \rangle  & = &
 i 2 \sqrt{m_{J/\psi} N_c} \epsilon_\mu \psi( 0).
\end{eqnarray}
In the above, $\epsilon_\mu$ is the polarization vector and $\psi(0)$
reduces to  the wave function of $J/\psi$ at origin in
the non-relativistic limit. Defining the charm quark pair polarization
function $\Pi(q)=-\Pi_\mu^\mu(q)/(3q^2)$, we find the $J/\psi$
contribution to its  imaginary part to be
\begin{eqnarray}
\frac{1}{\pi} {\rm Im} \Pi(s)= f_0 \delta(s-m_{J/\psi}^2),
\label{delta}
\end{eqnarray}
where
\begin{eqnarray}
f_0=\frac{12 \pi }{ m_{J/\psi}} |\psi( 0)|^2.\label{f0}
\end{eqnarray}
This quantity is related to the dielectron decay width of $J/\psi$, $
\Gamma_{J/\psi}^{e^+e^-}=\frac{16\pi\alpha^2 e_Q^2}{m_{J/\psi}^2}~|\psi(0)|^2$. Using the empirical value
$\Gamma_{J/\psi}^{e^+e^-}=$ 5.55 keV for $J/\psi$ in vacuum \cite{pdg}, we obtain the value
\begin{eqnarray}
|\psi(0)|=0.211 \pm 0.04 ~{\rm GeV}^{3/2} \label{overlap}
\end{eqnarray}
for the overlap of the charm vector current with a free $J/\psi$ or the
wave function of a free $J//\psi$ at origin. We note that in identifying $\psi(0)$
as the $J/\psi$ non-relativistic wave function at origin, Eq.~(\ref{f0}) is
expected to have several non-trivial corrections~\cite{Voloshin:2007dx}.
For example, within the non-relativistic limit, we could have  used
$2m_c$, with $m_c$ being the charm quark mass, in the denominator,
and this would amount to about 20\% correction and hence the uncertainty
given in Eq.~(\ref{overlap}).  Since the relativistic corrections are
expected to be temperature independent, we will study the temperature
dependence of the overlap of charm vector current with
$J/\psi$ at finite temperature relative to its vacuum value.

The residue of the charmonium correlator $f_0$ given in Eq.~(\ref{f0})
can be reliably calculated in the QCD sum rule
method~\cite{Shifman:1978bx,RRY}.   While masses of heavy
quark systems have been mostly studied in the moment sum rule, we use
here the Borel transformed sum rule, which suppresses the contribution
from the continuum, to calculate $f_0$ as it is more
robust under changes of the continuum~\cite{Bertlmann:1981he}.
The generalization of the Borel
sum rule to finite temperature is well founded as the operator product
expansion (OPE) is well known up to dimension 4 operators, which are
needed for the present analysis, by using the lattice data on the energy
density and pressure of hot QCD matter.  The method has been used to study the masses of
$J/\psi$ at finite density~\cite{Klingl:1998sr} as well as at finite
temperature~\cite{Morita:2007pt}.  Recently, the method has been
combined with the maximum entropy method to reconstruct the charmonium
spectral density at finite temperature~\cite{Gubler:2011ua}.  The Borel sum rule for $f_0$ only involves
the strength at the pole, which is an integrated quantity, and thus
does not depend on the details of the spectral density.  Hence, we concentrate on the temperature dependence of $f_0$.

Specifically, the OPE for the charm quark pair polarization function
$\Pi(q)$  is  equated to the spectral density
via the following dispersion relation after Borel transformation:
\begin{eqnarray}
{\cal M}(M^2)=\int_0^\infty ds e^{-s/M^2}  {\rm Im} \Pi(s),
\end{eqnarray}
after neglecting the thermal factor $\tanh[{s/2T}]$ in the spectral density as its correction to charmonium or the continuum is negligible up to temperatures of $1.1 T_c$~\cite{Morita:2007pt,Morita:2007hv}.
For the spectral density, we assume the following form:
\begin{eqnarray}
{\rm Im} \Pi(s)= {\rm Im} \Pi^{J/\psi}(s) + \theta(s-s_0) {\rm Im} \Pi^{\rm pert}(s) \label{Im}
\end{eqnarray}
with the first term given by the pole form in
Eq.~(\ref{delta}).

The Borel transformed OPE and its temperature dependence  up to dimension 4 operators are:
\begin{eqnarray}
{\cal M} (M^2) & = & e^{-\nu} \pi A(\nu) [1+ \alpha_s(M^2) a(\nu) +b(\nu) \phi_b(T) \nonumber\label{eq:OPE} \\
& & + c(\nu) \phi_c(T) ]
\end{eqnarray}
with $\nu=4m_c^2/M^2$ being a dimensionless scale parameter. The Wilson
coefficients $a(\nu)$, $b(\nu)$, and $c(\nu)$ are summarized in
Ref.~\cite{Morita:2009qk}, where the temperature dependence of the
scalar and twist-2 gluon condensates $\phi_b$ and $\phi_c$ are also
given. We note that the truncation of the OPE up to dimension 4 operators is valid up to temperature of $T=1.1T_c$ as well~\cite{Morita:2007hv}.

\begin{figure}[h]
\centerline{\includegraphics[width=8 cm]{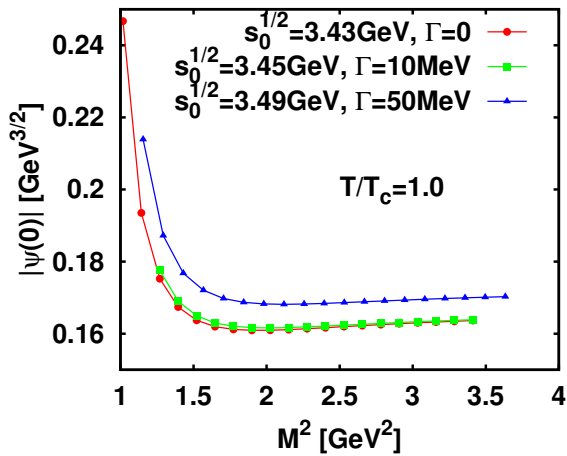}}
\caption{(Color online) Borel curves for $|\psi(0)|$ at $T=1.0~T_c$ for different values of $J/\psi$
 width and continuum threshold.}
\label{borel-stability}
\end{figure}

Using the OPE side of Eq.\eqref{eq:OPE}, we
can express the residue $f_0$ as
\begin{equation}
 f_0 = e^{m_{J/\psi}^2(M^2)/M^2}[\mathcal{M}^{\text{OPE}}(M^2)-\mathcal{M}^{\text{cont}}(M^2;s_0)],
\end{equation}
where the $J/\psi$ mass is given by
\begin{equation}
m_{J/\psi}^2 = -\frac{\frac{\partial}{\partial(1/M^2)}(\mathcal{M}^{\text{OPE}}(M^2)
 -\mathcal{M}^\text{cont}(M^2;s_0))}{\mathcal{M}^{\text{OPE}}(M^2)
 -\mathcal{M}^\text{cont}(M^2;s_0)}. \label{sumrule-mass}
\end{equation}
The continuum part of the correlator $\mathcal{M}^{\text{cont}}$ depends
on the threshold parameter $s_0$. Following the method used in
Ref.~\cite{Morita:2012m0m2} for the estimation of  $m_{J/\psi}$, we
determine $s_0$ by requiring  $|\psi(0)|=m_{J/\psi}f_0/(12\pi)$ to be
least sensitive to $M^2$. We have confirmed that the $M^2$ dependence
of $|\psi(0)|$ is quite similar to that of $m_{J/\psi}$, but the
resultant value of $s_0$  differs slightly from the case in which the
Borel curve for $m_{J/\psi}$ is optimized. We have also estimated  the
systematic error due to the $M^2$ dependence of $|\psi(0)|$ and found it
to be at most 0.001~GeV$^{3/2}$. We have further examined the effects of
finite width above $T_c$. Introducing the width, however, brings an
ambiguity on the stabilization of the Borel curve due to possible
absence of a solution \cite{Morita:2009qk}. Nevertheless, the change
due to the introduction of nonzero width can be understood as a result of the strong
correlation among the mass, the width and the threshold as shown in
Ref.~\cite{Morita:2009qk}. Since a nonzero width leads to a larger mass while keeping
the residue unchanged, the resultant $|\psi(0)|$ becomes slightly larger
as shown in Fig.~\ref{borel-stability} by the Borel curves for
$|\psi(0)|$ at $T=1.0~ T_c$ for three different values of the $J/\psi$
width.   Nevertheless, the difference is small and our prediction for $|\psi(0)|$ is thus robust.

\begin{figure}[h]
\centerline{
\includegraphics[width=8 cm]{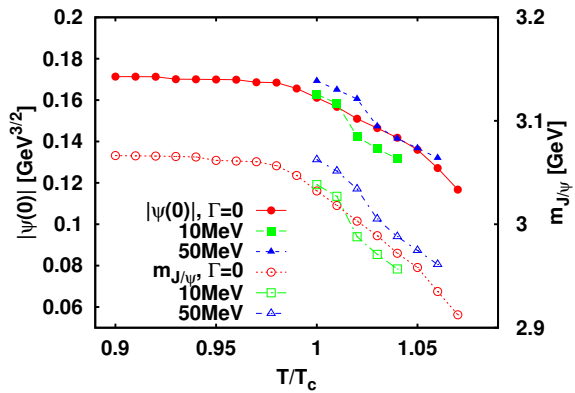}}
\caption{(Color online) Temperature dependence of $|\psi(0)|$ (filled symbols, left vertical axis) and mass of
$J/\psi$ (open symbols, right vertical axis) obtained from the QCD sum rule for different values of $J/\psi$ width.}
\label{temperature-dependence}
\end{figure}

Figure \ref{temperature-dependence} shows the temperature dependence of
$|\psi(0)|$ and $m_{J/\psi}$ obtained by the Borel stability analysis at each temperature
up to slightly above $T=1.05~ T_c$ beyond which the OPE becomes less
reliable~\cite{Morita:2007pt}. Irrespective of the $J/\psi$ width 
used in the analysis, both $|\psi(0)|$ (filled symbols, left vertical axis) and $m_{J/\psi}$
(open symbols, right vertical axis) are seen to decrease with increasing temperature.

To find out which potential between a charm and anticharm quark pair
correctly reproduces  $|\psi(0)|$
obtained from the QCD sum rule, we solve the following Schr\"{o}dinger
equation between a charm and anticharm pair:
\begin{eqnarray}
\bigg[2m_c-\frac{1}{m_c}\nabla^2+V(r,T)\bigg]\psi(r,T)=M \psi(r,T),\label{schrodinger1}
\end{eqnarray}
where $m_c=1.25$ GeV is the bare mass of charm  quark  and $\psi(r,T)$
is the charmonium wave function at temperature $T$.
Introducing the potential energy at infinitely large distance,
$V(r=\infty, T)$, the Schr\"odinger equation is modified
to~\cite{Karsch:1987pv,Satz:2005hx}
\begin{align}
\bigg[-\frac{\nabla^2}{m_c}+\widetilde{V}(r,T)\bigg]\psi(r,T)
=-\varepsilon\psi(r,T),
\label{schrodinger2}
\end{align}
where $\widetilde{V}(r,T)\equiv V(r,T)-V(r=\infty, T)$, and it vanishes
at infinitely large distance,  and
$\varepsilon=2m_c+V(r=\infty,
T)- M$ is the binding energy of  $J/\psi$ at temperature $T$. For
the heavy quark potential, we use either the free
energy between a heavy quark-antiquark pair that is extracted
from lattice calculations~\cite{Digal:2005ht,Satz:2005hx,Kaczmarek:1900zz}
or the more attractive one based on the internal energy by adding the
contribution from the entropy density.

\begin{figure}[h]
\centerline{
\includegraphics[width=8 cm]{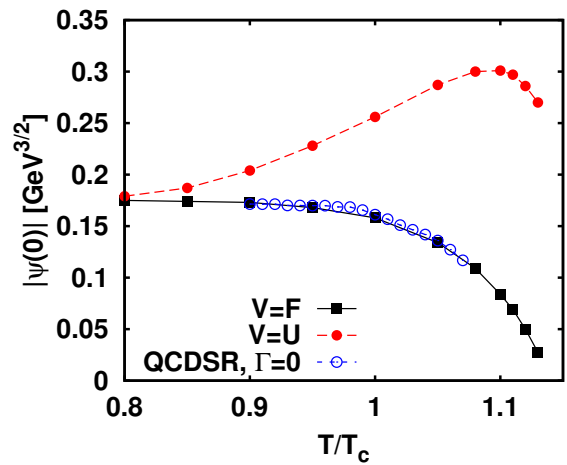}}
\caption{(Color online) Temperature dependence of $|\psi(0)|$ obtained from the free energy (filled squares) and internal energy (filled circles) potentials together with that from the QCD sum rule (open circles).}
\label{potential}
\end{figure}

Figure \ref{potential} shows the temperature dependence of the strength
of the $J/\psi$ wave function at the origin obtained by solving the
Schr\"{o}dinger with the two different potentials, together with the result
from the QCD sum rule in the case of vanishing width.  With the internal energy as the potential, the strength is seen to increase by almost a factor of two
at slightly above the critical temperature.
On the other hand, the strength decreases
monotonically with temperature when the free energy is used as the potential,
strikingly similar to the result from QCD sum rules in both its behavior and  values.
Even allowing for $\pm 20$\% uncertainty in the sum
rule result, the case with internal energy is outside the range of
expectations from sum rule calculations. Hence we can conclude that to
correctly reproduce the non-relativistic wave function of $J/\psi$ at
finite temperature, one should use the free energy potential in the
Schr\"{o}dinger equation.

\begin{figure}[h]
\centerline{\includegraphics[width=8 cm]{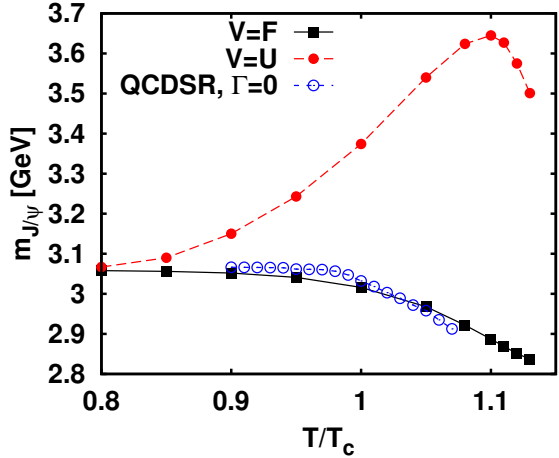}}
\caption{(Color online) Temperature dependence of $m_{J/\psi}$ obtained from solving the Schr\"{o}dinger equation using either the free energy (filled squares) or the internal energy (filled circles)  and from the QCD sum rule analysis (open circles).}
\label{mass}
\end{figure}

In Fig.~\ref{mass}, we plot the mass of $J/\psi$ obtained from
Eq.~(\ref{schrodinger1}) with different potentials, and compare the
results to that obtained from the sum rule analysis of
Eq.~(\ref{sumrule-mass}).  Again, we find that the result obtained from
using the free energy is consistent with that from the QCD sum rule
analysis.

The result obtained in the present study that the potential between a heavy quark-antiquark pair in QGP is determined by their free energy has important phenomenological consequences.   For $J/\psi$, this leads to a dissociation temperature that is only slightly above $T_c$~\cite{Kaczmarek:2005uv}. As a result,  all  charmonium states would essentially melt in the quark-gluon plasma. The observed $J/\psi$'s in relativistic heavy ion collisions are then either from initial hard collisions that occur outside the QGP or from the recombination of charm and anticharm quarks in the QGP during hadronization.

It should be noted, however, that in obtaining our result from the Schr\"odinger equation, we have kept the charm quark mass  to its zero temperature value and also taken  the heavy quark potential to be real. These quantities taken separately are  gauge dependent, and different prescriptions can be used to determine their values at finite temperature.  Since the heavy quark potential is expected to acquire a gauge dependent imaginary part at finite temperature~\cite{Laine:2006ns,Rothkopf:2011db},  a stronger potential and a different charm quark in-medium mass may then be needed to reproduce gauge invariant and physically observable quantities like $|\psi(0)|$, $m_{J/\psi}$ and the $J/\psi$ dissociation temperature.

\textit{Acknowledgements}
We thank useful comments from A. Rothkopf and Y. Akamatsu. This work was supported in part by the Korean Research Foundation under Grant Nos. KRF-2011-0020333 and KRF-2011-0030621, the Yukawa International Program for
Quark-Hadron Sciences at Kyoto University and the Grant-in-Aid for Scientific Research from JSPS No.~24540271 as well as the US National Science Foundation under Grant No. PHY-1068572, the US Department of
Energy under Contract No. DE-FG02-10ER41682, and the Welch Foundation
under Grant No. A-1358.

\end{document}